# A First-Principles Study on the Adsorption of Small Molecules on Antimonene: Oxidation Tendency and Stability

Andrey A. Kistanov,[a, b] Yongqing. Cai,*[, b] Devesh R. Kripalani,[a] Kun Zhou,*[, a] Sergey V. Dmitriev[c, d] and Yong-Wei Zhang*[b]

Antimonene, a new group-VA 2D semiconducting material beyond phosphorene, was recently synthesized through various approaches and was shown to exhibit a good structural integrity in ambient conditions and various interesting properties. In this work, we perform systematical first-principles investigations on the interactions of antimonene with the small molecules CO, NO, $NO_2$, $H_2O$, $O_2$, $NH_3$ and $H_2$. It is found that NO, $NO_2$, $H_2O$, $O_2$, and $NH_3$ serve as charge acceptors, while CO shows a negligible charge transfer. $H_2$ acts as a charge donor to antimonene with the amount of charge transfer being ten times that of $H_2$ on phosphorene. The interaction of the $O_2$ molecule with antimonene is much stronger than that with phosphorene. Surprisingly, the kinetic barrier for the splitting of the $O_2$ molecule on antimonene is low (~0.40 eV), suggesting that pristine antimonene may undergo oxidation in ambient conditions, especially at elevated temperatures. Fortunately, the acceptor role of $H_2O$ on antimonene, opposite to a donor role in phosphorene, helps to suppress further structural degradation of the oxidized antimonene by preventing the proton transfer between water molecules and oxygen species to form acids. By comparing antimonene with phosphorene and InSe, we suspect that the acceptor role of water may be a necessary condition for a good environmental stability of such 2D layers to avoid structural decomposition. While the surface oxidation layer may serve as an effective passivation layer from further degradation of the underlying layers, our findings show that antimonene layers still need to be separated or properly protected by other noncovalent functionalization from oxygen or other environmental molecules. The present work reveals interesting insights into the environmental effects of physisorbed small molecules on the oxidation tendency and stability of antimonene, which may be important for its growth, storage and applications.

**Introduction**

Two-dimensional (2D) materials, such as graphene, phosphorene, InSe, and transition metal dichalcogenides (TMDs), have attracted a great deal of attention in recent years due to their intriguing electronic, optical and other physical properties.[1-7] More recently, for the first time, antimonene, which is a monolayer of antimony, a new group-VA 2D semiconducting material, was predicted by density functional theory calculations.[8] Subsequently, a high-quality antimonene sheet was obtained by mechanical exfoliation of bulk antimony,[9,10] and large-sized few-layer antimonene was produced by liquid-phase exfoliation and epitaxy growth on various substrates like $SiO_2$, $Bi_2Te_3$, and $Sb_2Te_3$.[11-14] Importantly, antimonene exhibits a high mechanical stability[9] and unique thermodynamic stability in ambient conditions,[12] in strong contrast to the poor stability of phosphorene. Antimonene also possesses a buckled structure, a wide indirect band gap of 2.28 eV, and a high carrier mobility.[15-17] A previous theoretical study suggested that external strain can be used to tune antimonene from an indirect to a direct band gap semiconductor.[17] Therefore, due to its tunable wide band gap, anisotropic carrier mobility, excellent optical and thermoelectric response and high structural stability at ambient conditions,[2, 17-23] layered antimonene is promising for various potential applications, such as sensors,[22, 24-26] spintronics,[27, 28] energy storage and conversion devices.[29, 30]

Owing to their atomically thin structures, high surface−volume ratio and weak electronic screening, 2D materials, such as graphene, phosphorene, InSe and $MoS_2$, tend to be sensitive to the exposure of external adsorbates including environmental molecules and dopants.[24,31] For example, phosphorene undergoes structural degradation upon exposure to oxygen and water molecules.[32-36] External molecules and dopants can also enhance electronic properties and chemical activities of 2D materials by donating electrons/holes or by altering the work function of the host material.[25,26,37] For instance, it is well known that hydrogenation of graphene[38] is able to lead to a semimetal-to-insulator transformation, while surface patterning of $MoS_2$ with hydrogen may provide an effective way to create a metallic nanoroad for interconnection.[39] Selective surface decoration by molecules such as NO, $NO_2$ and $O_2$, and typical charge-transfer organic molecules, such as tetracyanoethylene, tetrathiafulvalene and tetracyanoquinodimethane,[40-43] was shown to cause alteration of carrier density, shift of the Fermi level and even change in the optical properties of many 2D materials.[33, 44-46] For example, their interactions were shown to cause the changes in excitons and photoluminescence peaks in TMDs[47] and lead to the development of phosphorene-based high-performance gas sensors.[48, 49] To apply the emerging layered antimonene for nanoelectronic and chemical applications, a comprehensive understanding of its interaction with many common environmental molecules is highly desired. However, effects of environmental molecules on the carrier density in antimonene and tendency of charge flow across their interfaces are still unknown.

In this work, by using first-principles calculations, we study the effects of physisorption of several small molecules, including CO, NO, $NO_2$, $H_2O$, $O_2$, $NH_3$ and $H_2$, on the electronic properties of monolayer antimonene. A thorough analysis on the charge transfer across molecular adsorbate-antimonene interfaces is carried out. In particular, the stability issue of antimonene under the environmental oxygen and water molecules is examined and discussed from the atomic scale. This topic has not been discussed before and is critically important for the synthesis, storage, and applications of antimonene.

**Computational details**

Our calculations are performed within the framework of the density functional theory (DFT) by using Vienna *ab initio* simulation package (VASP).[50] For a proper treatment of the noncovalent chemical interactions between antimonene and small molecules, a van der Waals-corrected functional with Becke88 optimization (optB88) is used.[51] All the considered structures (pure and antimonene adsorbed with molecules) are fully relaxed until the forces on each atom are smaller than 0.01 eV Å$^{-1}$. The relaxed lattice constants of monolayer antimonene are $a = b = 4.308$ Å and the calculated band gap is 1.14 eV (GGA method), which is consistent with the results of recent works.[28, 52, 53] To consider the effects of molecular adsorbates in the dilute doping limit, we place the molecule on an antimonene sheet consisting of a $4 \times 4 \times 1$ supercell (32 Sb atoms). To avoid the interaction between the replicate units, a vacuum space of 20 Å is applied. The first Brillouin zone is sampled with a $6 \times 6 \times 1$ k-mesh grid and a kinetic energy cutoff of 450 eV is adopted. The adsorption energy $E_a$ of a molecule on antimonene is calculated as $E_a = E_{Sb+mol} - E_{Sb} - E_{mol}$, where $E_{Sb+mol}$, $E_{Sb}$, and $E_{mol}$ are the energies of the molecule-adsorbed antimonene, the isolated antimonene, and the molecule, respectively. Charge transfer analysis is conducted by the calculation of the differential charge density (DCD) $\Delta\rho(r)$, which is defined as $\Delta\rho(r) = \rho_{Sb+mol}(r) - \rho_{Sb}(r) - \rho_{mol}(r)$, where $\rho_{Sb+mol}(r)$, $\rho_{Sb}(r)$, and $\rho_{mol}(r)$ are the charge densities of the molecule-adsorbed antimonene, the isolated antimonene, and the molecule, respectively. The exact amount of the charge transfer between the molecule and the surface is calculated by integrating $\Delta\rho(r)$ over the basal plane at the $z$ point for deriving the plane-averaged DCD $\Delta\rho(z)$ along the normal direction $z$ of the sheet. The amount of transferred charge at the $z$ point is given by $\Delta Q(z) = \int_{-\infty}^{z} \Delta\rho(z')dz'$. Based on the $\Delta Q(z)$ curve, the total number of electrons donated by the molecule is read at the interface between the molecule and antimonene where $\Delta\rho(z)$ shows a zero value.

**Results and discussion**

We consider the influence of small molecules CO, NO, $NO_2$, $H_2O$, $O_2$, $NH_3$ and $H_2$ on the electronic properties and chemical activities of antimonene. The adsorption energy and charge transfer between these molecules and the antimonene surface are systematically investigated. For each molecule, several different configurations and possible adsorbing sites are considered, including the top of the Sb site, the top site above the center of the hexagon, and the top site above the Sb-Sb bond with the molecules being aligned tilted, parallel or perpendicular to the surface. We test the random adsorption inside the hexagon, and find that those adsorbing configurations become similar to one of the mentioned three cases, but with slightly higher adsorption energy. The results of the adsorption energy $E_a$, the charge transfer $\Delta q$ and the shortest distance $d$ from the molecule to the Sb atom for the lowest-energy configuration are summarized in Table 1.

*CO adsorption.* The most stable configuration and the DCD isosurface plot for the CO molecule adsorbed on antimonene are shown in Figure 1a. The molecule adopts a tilted configuration above the center of the hexagon with $d = $ ~3.72 Å and $E_a = -0.12$ eV. The DCD isosurface plot (Figure 1a) clearly reflects an accumulation of electrons in the region around the C atom, indicating a loss of electrons in the proximity of the antimonene sheet. This is not surprising since elemental C is more electronegative than Sb. Quantitative DCD $\Delta\rho$(r) analysis (Figure 1b) reveals that only a tiny number of electrons are transferred from antimonene to the CO molecule (−0.003 $e$ per molecule), which is consistent with the almost unchanged C−O bond length compared with that of the isolated gas molecule. Interestingly, the values of $d$ and $E_a$, and the charge transfer ability of the CO molecule adsorbed on antimonene are similar to those of the CO molecule on InSe,[54] while significantly different from those for the CO molecule adsorbed on graphene[26] and phosphorene.[33, 55] This similarity of the CO molecule behavior on antimonene and InSe can be attributed to their similar honeycomb structure with the lone-pair electrons associated with Sb or Se atoms. The band structure of antimonene adsorbed with the CO molecule (Figure 1c) clearly shows that there are no additional CO-induced states within the band gap of antimonene. The value of the band gap is almost unchanged compared with that of pristine antimonene (1.14 eV). The local density of states (LDOS) plot (Figure 1d) shows that the highest occupied molecular orbital (HOMO) 5σ and the lowest unoccupied molecular orbital (LUMO) $2\pi^*$ of the CO molecule adsorbed on antimonene are located at −4.30 and 2.10 eV (relative to the Fermi level), respectively. It should be noted that the HOMO level is a non-resonant state located below the valence band of antimonene, while the LUMO level is located within the conduction band. The $2\pi^*$ peak is significantly broadened compared with the 5σ level, which is opposite to the case of CO above InSe.[54] This alignment of CO LUMO states within the conduction band of antimonene suggests that the photo-excited electrons of antimonene may partially transfer to the CO LUMO state, which can trigger a different electron-hole recombination rate and prolong the lifetime of holes in the antimonene sheet upon exposure to the CO gas. On the other hand, the enhanced occupation of this antibonding orbital should weaken the C-O bond and affect its infrared frequency, which would allow the monitoring of the population of photo-excited carriers in antimonene.

**Table 1.** Adsorption energy $E_a$, the amount of charge transfer $\Delta q$, the shortest distance $d$ from the molecule to the Sb atom, and the donor/acceptor characteristics of the molecular dopant on the antimonene surface. Note that a positive (negative) $\Delta q$ indicates a loss (gain) of electrons from each molecule to antimonene.

| Molecule | Antimonene | | | | Phosphorene [Ref. 33] | | InSe [Ref. 54] | |
|---|---|---|---|---|---|---|---|---|
| | $d$ (Å) | Doping nature | $E_a$ (eV) | $\Delta q$ (e) | $E_a$ (eV) | $\Delta q$ (e) | $E_a$ (eV) | $\Delta q$ (e) |
| CO | 3.72 | – | -0.12 | -0.003 | -0.31 | 0.007 | -0.13 | 0.001 |
| NO | 2.70 | acceptor | -0.44 | -0.067 | -0.32 | -0.074 | -0.13 | -0.094 |
| $NO_2$ | 2.44 | acceptor | -0.81 | -0.156 | -0.50 | -0.185 | -0.24 | -0.039 |
| $H_2O$ | 2.98 | acceptor | -0.20 | -0.021 | -0.14 | 0.035 | -0.17 | -0.01 |
| $O_2$ | 3.21 | acceptor | -0.61 | -0.116 | -0.27 | -0.064 | -0.12 | -0.001 |
| $NH_3$ | 3.41 | acceptor | -0.12 | -0.029 | -0.18 | 0.050 | -0.20 | -0.019 |
| $H_2$ | 3.56 | donor | -0.04 | 0.138 | -0.13 | 0.013 | -0.05 | 0.146 |

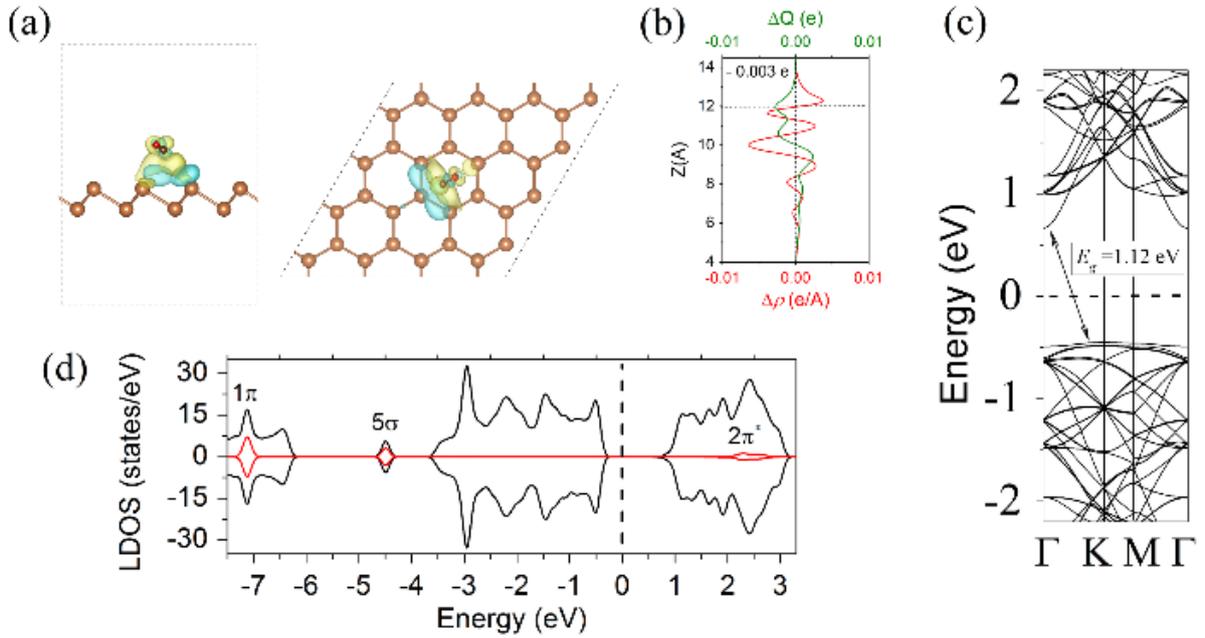

**Figure 1.** (a) The top and side views of the lowest-energy configuration combined with the DCD isosurface plots ($10^{-3}$ Å$^{-3}$) for antimonene adsorbed with the CO molecule. The yellow (blue) color represents an accumulation (depletion) of electrons. (b) The line profiles of the plane-averaged $\Delta\rho(z)$ (red line) and the transferred amount of charge $\Delta Q(z)$ (green line). (c) The band structure of antimonene adsorbed with the CO molecule. The black dashed lines show the Fermi level. (d) The total DOS (black line) and LDOS (red line) of antimonene adsorbed with the CO molecule. The black dashed lines show the Fermi level.

***NO adsorption.*** Figure 2a shows the most stable configuration and the DCD isosurface plot for the NO molecule adsorbed on antimonene. Similar to the CO adsorption, NO adopts a tilted configuration and is located above the center of the hexagon with $d = \sim 2.70$ Å. However, as a typical open-shell molecule, NO has a much stronger interaction with the underlying antimonene with $E_a = -0.44$ eV. The magnitude is slightly larger than that of NO on phosphorene ($-0.32$ eV).[33] The DCD plot in Figure 2a depicts the orbital-like lobes of the diminishing and accumulating electronic densities, which suggest a redistribution of the surface charges of antimonene upon the NO adsorption. On the other hand, the population of some NO molecule orbitals becomes less occupied upon interaction with antimonene, while for antimonene most of the transferred electrons are distributed at the Sb atoms closest to the NO molecule. The charge transfer analysis (Figure 2b) reveals that NO acts as a strong acceptor to antimonene with a charge transfer of $-0.067$ $e$, which is similar to the role of NO on phosphorene[33, 55] and InSe.[54]

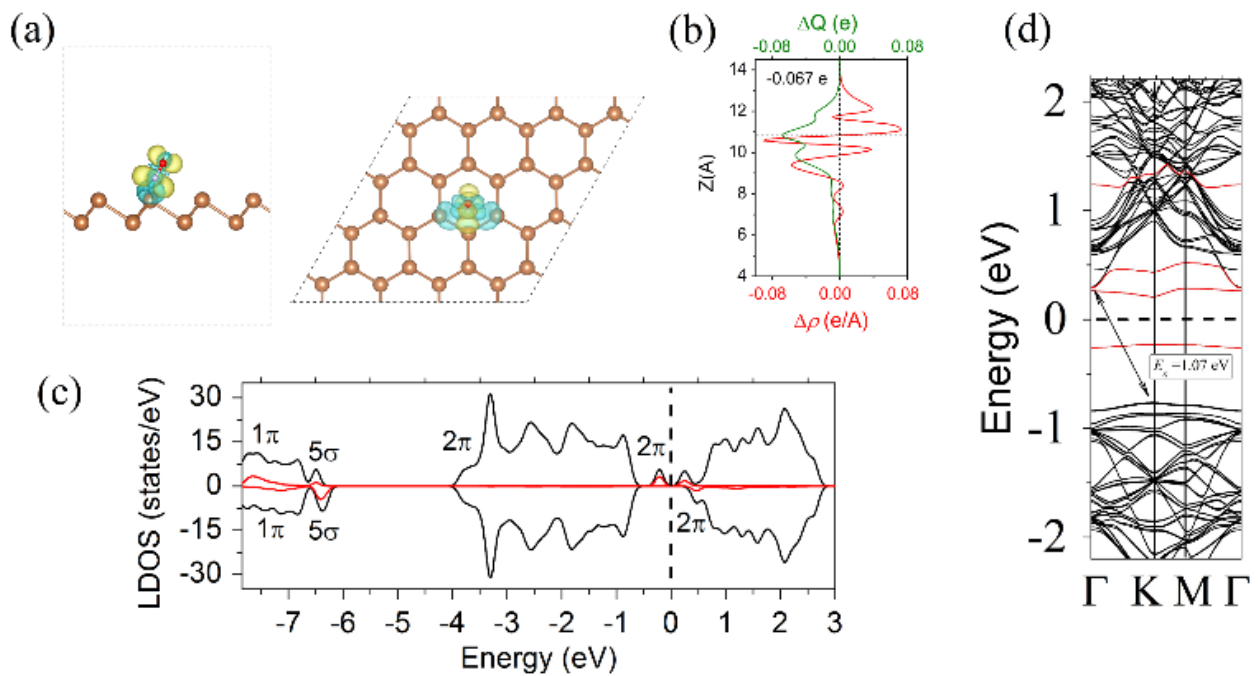

**Figure 2.** (a) The top and side views of the lowest-energy configuration combined with the DCD isosurface plots (0.001 Å$^{-3}$) for antimonene adsorbed with the NO molecule. The yellow (blue) color represents an accumulation (depletion) of electrons. (b) The line profiles of the $\Delta\rho(z)$ (red line) and the transferred amount of charge $\Delta Q(z)$ (green line). (c) The total DOS (black line) and LDOS (red line) of antimonene adsorbed with the NO molecule. The black dashed lines show the Fermi level. (d) The band structure of antimonene adsorbed with the

NO molecule. The bands colored in black and red represent antimonene and NO, respectively. The black dashed lines show the Fermi level.

The LDOS and band structure plots, shown in Figures 2c and d, respectively, reveal that the state hybridization and charge transfer between NO and antimonene lead to broadening and splitting of the degeneracy of NO orbitals. More specifically, the degeneracy of the $2\pi$ orbital is lifted and evolves into two levels located close to the conduction band minimum of the antimonene (see Figure 2c). In addition, the NO molecule level is spin-split, which induces a magnetic moment of 1 $\mu$B in the adsorbed system. The band gap of antimonene adsorbed with the NO molecule slightly decreases from 1.14 eV of the pristine sheet to 1.07 eV. The presence of NO-induced states within the band gap of antimonene (Figure 2d) can modify the optical and electronic properties of antimonene with the NO molecule serving as an electron trapping center.

*NO$_2$ adsorption.* NO$_2$ was predicted to have the strongest interaction among the typical small molecules in the cases of phosphorene[33] and InSe.[54] Such strong interaction was later demonstrated in experiments with the fabrication of phosphorene−based sensors, which show a high selectivity to the NO$_2$ gas in the presence of other gases.[48, 49] Here, for the NO$_2$ molecule adsorbed on antimonene, we predict a much stronger interaction ($E_a = -0.81$ eV) than on phosphorene ($E_a = -0.50$ eV[33]). The most stable configuration and the DCD isosurface plot for the NO$_2$ molecule adsorbed on antimonene are presented in Figure 3a. The molecule takes the position above the Sb-Sb bond with the two O atoms situated closer to the surface plane with $d = \sim 2.44$ Å. The N−O bond length ranges from 1.25 to 1.27 Å, which is slightly larger than the N−O bond length (1.20 Å) of the free NO$_2$ gas molecule, resulting from the strong molecule-antimonene interaction. The isosurface plot (Figure 3a) and the DCD analysis (Figure 3b) suggest a large electron transfer of −0.156 $e$ per molecule from the antimonene surface to the NO$_2$.

The LDOS plot (Figure 3c) shows that the 6a$_1$ orbital is split into two levels located within the band gap of antimonene, the LUMO (6a$_1$, spin-down) just above and the HOMO state (6a$_1$, spin-up) just below the Fermi level, leading to a magnetic moment of 1 $\mu$B. In addition, the 4b$_1$ and 1a$_2$ NO$_2$ orbitals significantly broaden and coincide with the valence states of antimonene, while the 5b$_1$ state coincides with the conduction states of antimonene. Such orbital mixing and hybridization can facilitate the charge transfer between NO$_2$ and antimonene. The presence of the NO$_2$ molecule induces localized states within the band gap

(Figure 3d), which may affect the optical properties of antimonene. On the other hand, the change of the band gap of the host antimonene (1.17 eV) is negligible compared with that of pristine antimonene (1.14 eV).

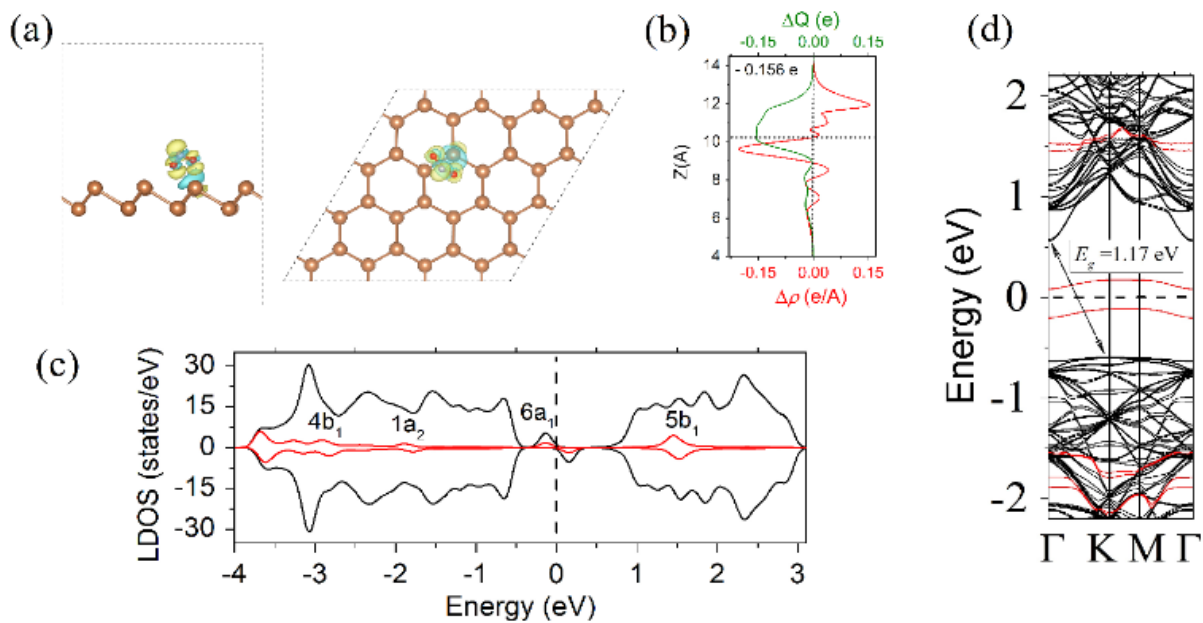

**Figure 3.** (a) The top and side views of the lowest-energy configuration combined with the DCD isosurface plots (0.003 Å$^{-3}$) for antimonene adsorbed with the $NO_2$ molecule. The yellow (blue) color represents an accumulation (depletion) of electrons. (b) The line profiles of the $\Delta\rho(z)$ (red line) and the transferred amount of charge $\Delta Q(z)$ (green line). (c) The total DOS (black line) and LDOS (red line) of antimonene adsorbed with the $NO_2$ molecule. The black dashed lines show the Fermi level. (d) The band structure of antimonene adsorbed with the $NO_2$ molecule. The bands colored in black and red represent antimonene and $NO_2$, respectively. The black dashed lines show the Fermi level.

***$H_2O$ and $O_2$ adsorptions.*** Effects of $H_2O$ and $O_2$ molecules on the electronic properties and charge transfer of 2D materials are highly important with regard to the carrier density and structural stability. The most stable configuration and the DCD isosurface plot for the $H_2O$ and $O_2$ molecules adsorbed on antimonene are presented in Figures 4a and 5a, respectively. The $H_2O$ molecule adopts a flat alignment relative to the antimonene basal plane and is located at $d = \sim2.98$ Å. The $O_2$ molecule adopts a tilted configuration and is located at the center of the hexagon at $d = \sim3.21$ Å. The $H_2O$ molecule possesses a relatively weak $E_a = -0.20$ eV, while the $O_2$ molecule has a much larger $E_a = -0.61$ eV.

For the $H_2O$ molecule, the DCD plot (Figure 4a) together with the charge transfer analysis (Figure 4b) show an accumulation of electrons in the $H_2O$ molecule (acceptor to antimonene) with a total charge transfer of ~−0.021 $e$ per molecule. The $O_2$ molecule also acts as an acceptor with the total transferred charge of ~ −0.116 $e$ per molecule (Figure 5b). The bond length of the adsorbed $O_2$ molecule is 1.26 Å, comparable to 1.22 Å of the free molecule. Therefore, similar to phosphorene, antimonene shows high oxidation ability and may oxidize easily at ambient conditions. On the other hand, antimonene demonstrates a weaker interaction with the $H_2O$ molecule, which is similar to phosphorene[32] and graphene.[56] Our calculation suggests that the main source of antimonene oxidation originates from the presence of $O_2$ rather than $H_2O$, owing to a stronger binding strength and a larger amount of charge transfer of $O_2$.

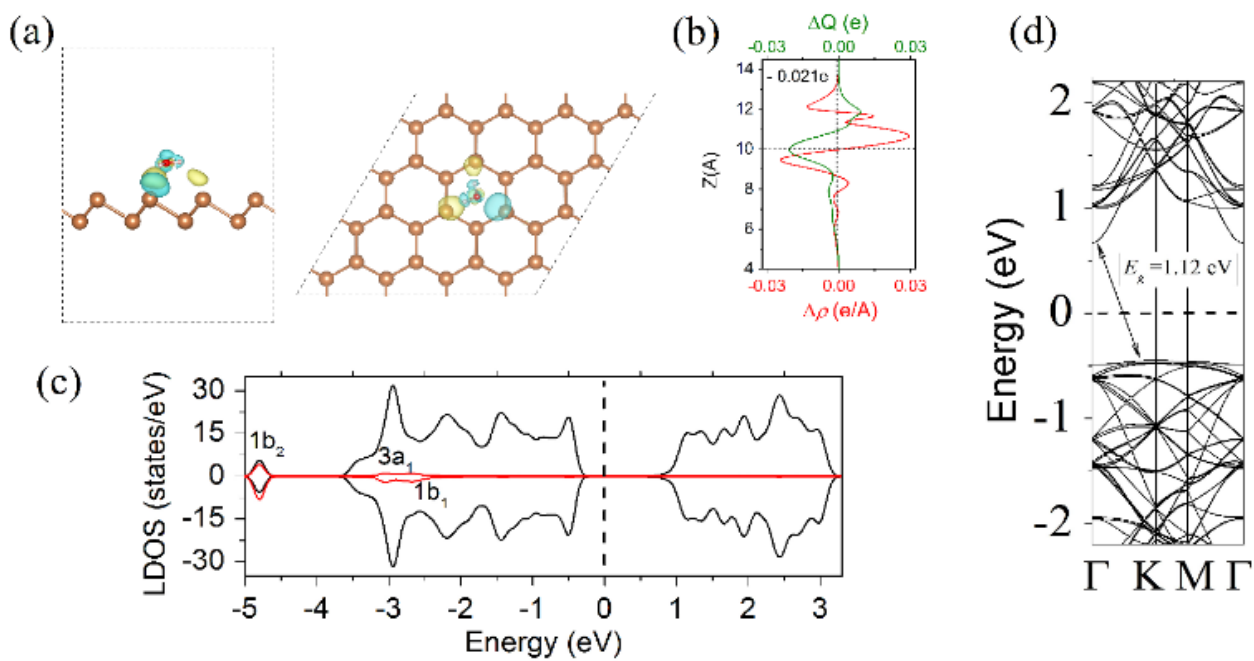

**Figure 4.** (a) The top and side views of the lowest-energy configuration combined with the DCD isosurface plots (0.6·10$^{-3}$ Å$^{-3}$) for antimonene adsorbed with the $H_2O$ molecule. The yellow (blue) color represents an accumulation (depletion) of electrons. (b) The line profiles of the $\Delta\rho(z)$ (red line) and the transferred amount of charge $\Delta Q(z)$ (green line). (c) The total DOS (black line) and LDOS (red line) of antimonene adsorbed with the $H_2O$ molecule. The black dashed lines show the Fermi level. (d) The band structure of antimonene adsorbed with the $H_2O$ molecule. The black dashed lines show the Fermi level.

The LDOS and band structure plots for antimonene adsorbed with the $H_2O$ molecule (Figures 4c and d) indicate the absence of $H_2O$-induced states within the band gap of antimonene. In addition, the $1b_2$, $3a_1$, and $1b_1$ orbitals of the $H_2O$ molecule significantly broaden and coincide with the valence states of antimonene (Figure 4c). This indicates that the performance of antimonene, such as durability and carrier mobility, tends to be affected by the presence of moisture due to the strong state coupling. For antimonene adsorbed with the $O_2$ molecule, the LDOS and band structure (Figures 5c and d) reflect additional $O_2$-induced states within the band gap of antimonene. The Fermi level crosses the half-filled $2\pi$ HOMO state, which aligns slightly (~0.15 eV) above the valence band maximum, allowing the electrons to be excited to the $O_2$ molecule, and thereby creating holes in antimonene. The $2\pi^*$ LUMO state is located at 0.50 eV above the Fermi level (Figure 5c). The presence of the $O_2$-induced states within the band gap of antimonene and the non-trivial adsorption and oxidation ability of the $O_2$ molecule to antimonene can significantly alter the optical and electronic properties of antimonene. In addition, the band structure analysis also shows a small decrease of the band gap size from 1.14 eV of pristine antimonene to 1.08 eV ($O_2$-induced states are not taken into account) upon the $O_2$ molecule adsorption.

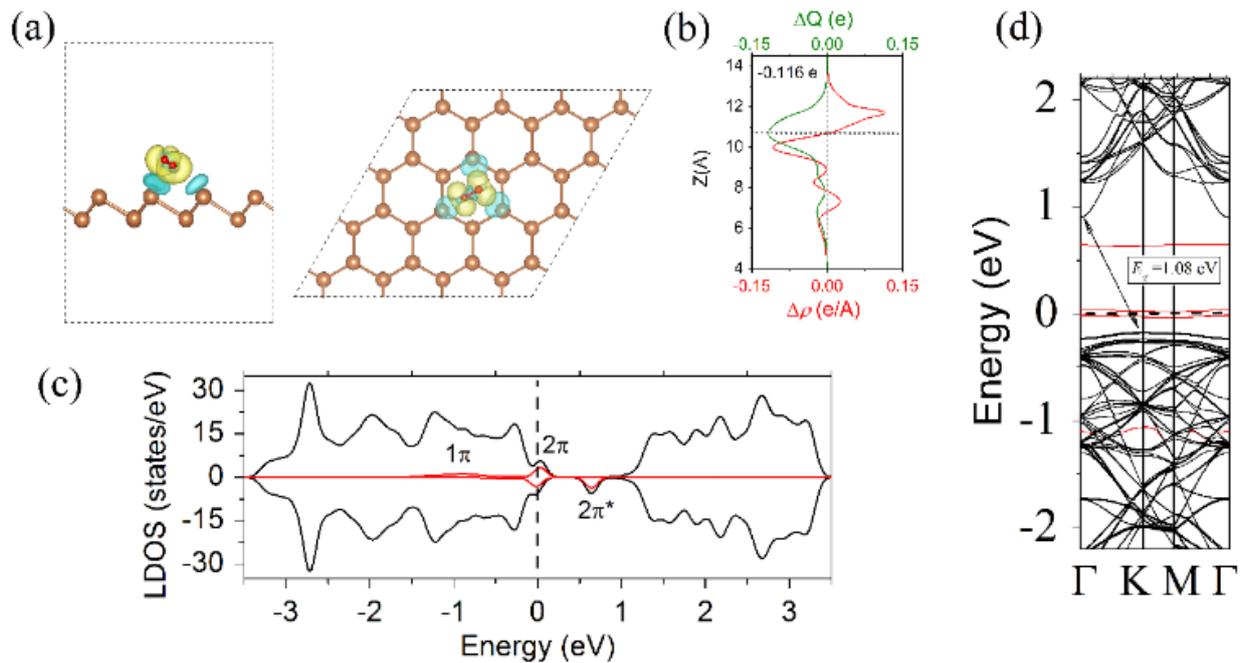

**Figure 5.** (a) The top and side views of the lowest-energy configuration combined with the differential charge density (DCD) isosurface plots ($0.7 \cdot 10^{-3}$ Å$^{-3}$) for antimonene adsorbed with the $O_2$ molecule. The yellow (blue) color represents an accumulation (depletion) of electrons. (b) The line profiles of the plane-averaged differential charge density $\Delta\rho(z)$ (red line) and the transferred amount of charge $\Delta Q(z)$ (green line). (c) Total DOS (black line) and LDOS (red line) of antimonene adsorbed with the $O_2$ molecule. The bands colored in black

and red represent antimonene and $O_2$, respectively. The black dashed lines show the Fermi level. (d) The band structure of antimonene adsorbed with the $O_2$ molecule. The black dashed lines show the Fermi level.

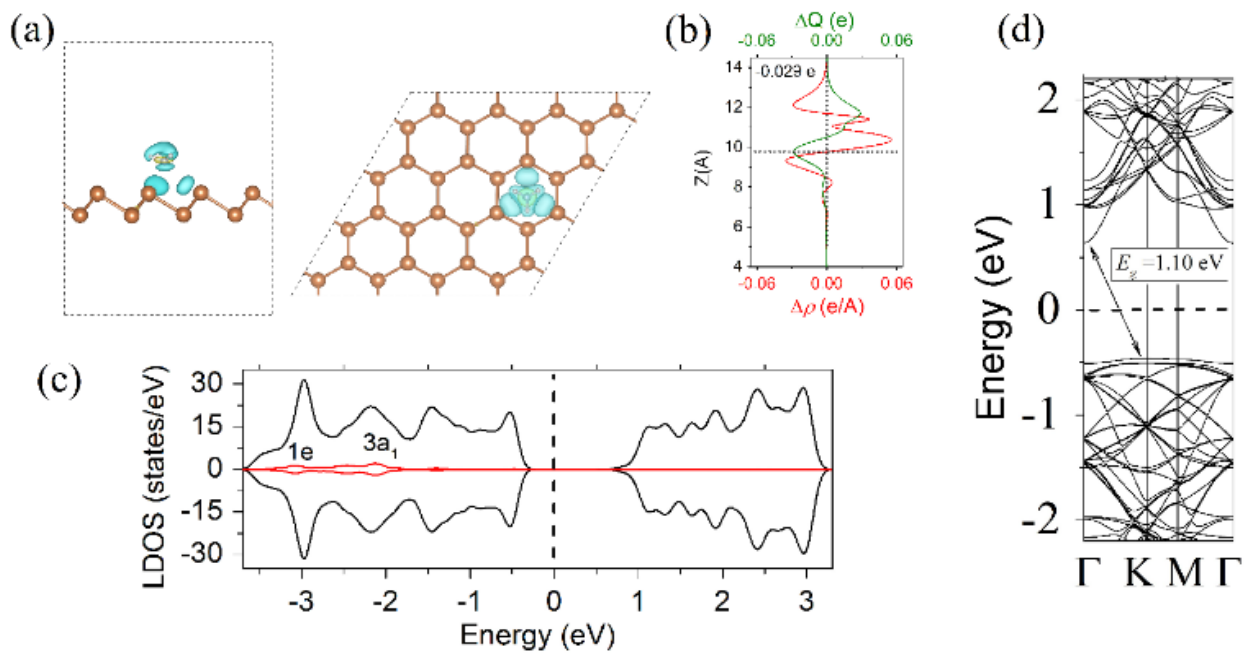

**Figure 6.** (a) The top and side views of the lowest-energy configuration combined with the DCD isosurface plots ($0.6 \cdot 10^{-3}$ Å$^{-3}$) for antimonene adsorbed with the $NH_3$ molecule. The yellow (blue) color represents an accumulation (depletion) of electrons. (b) The line profiles of the $\Delta\rho(z)$ (red line) and the transferred amount of charge $\Delta Q(z)$ (green line). (c) Total DOS (black line) and LDOS (red line) of antimonene adsorbed with the $NH_3$ molecule. The bands colored in black and red represent antimonene and $NH_3$, respectively. The black dashed lines show the Fermi level. (d) The band structure of antimonene adsorbed with the $NH_3$ molecule. The black dashed lines show the Fermi level.

***$NH_3$ adsorption.*** Concerning the adsorption of the $NH_3$ molecule, the lowest energy configuration is found when the molecule is located at $d = 3.41$ Å at the hollow hexagon center with the N atom pointing towards the surface and the three H atoms pointing away from the surface (Figure 6a). The $E_a$ is −0.12 eV and the lengths of the three N−H bonds are all 1.024 Å, which is comparable with the N−H bond length of 1.01 Å of the $NH_3$ gas molecule. Charge analysis (Figure 6a) shows that electrons are transferred to the $NH_3$ molecule from the nearest Sb atoms. The total charge transfer from the antimonene surface to the $NH_3$ molecule is found to be as high as −0.029 $e$ per molecule (Figure 6b). A similar acceptor behavior is predicted for the NH$_3$ molecule adsorbed on InSe[54], while for graphene[26] and phosphorene[33, 55], the NH$_3$ molecule acts as a donor. The underlying reason for the acceptor role of NH$_3$ is that the N atom is more electronegative than the Sb and Se atoms. The LDOS analysis (Figure 6c) shows that the nonbonding 3a$_1$ and the doubly degenerated 1e HOMO orbitals are significantly below the Fermi level and largely broadened, which can indicate the hybridization of these states with the valence states of antimonene. The band structure analysis (Figure 6d) reveals no significant change in the band gap size of antimonene adsorbed with the NH$_3$ molecule compared with that of pristine antimonene.

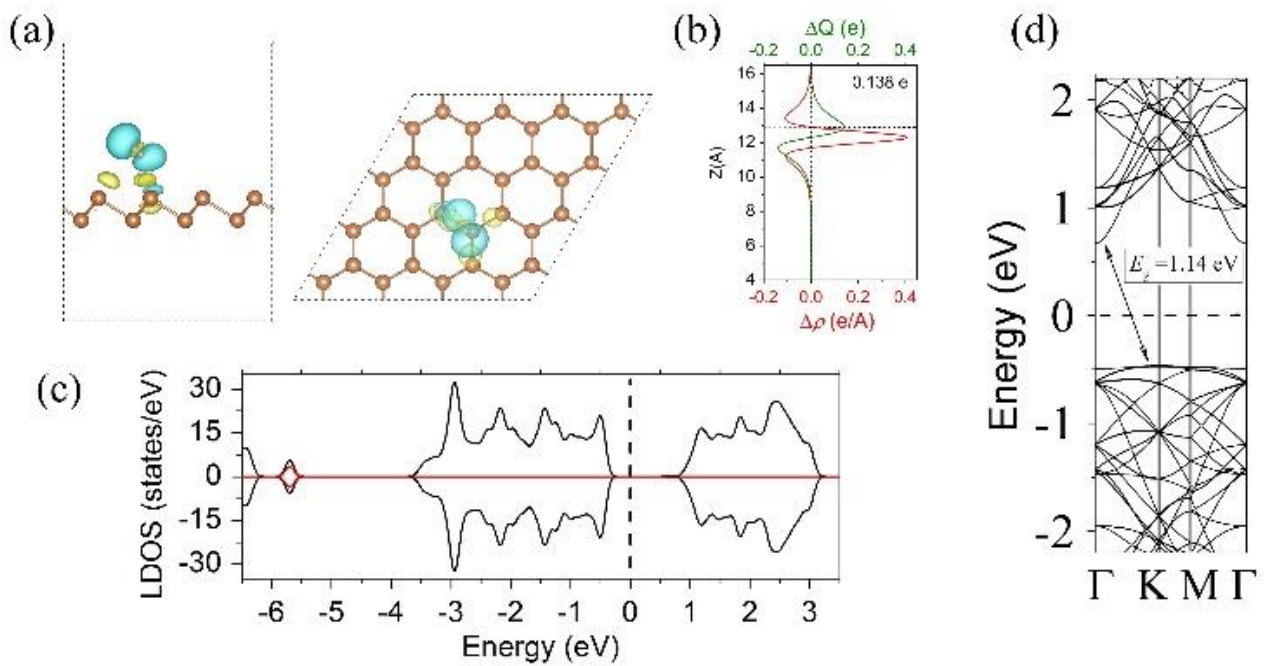

**Figure 7.** (a) The top and side views of the lowest-energy configuration combined with the DCD isosurface plots (0.6·10$^{-3}$ Å$^{-3}$) for antimonene adsorbed with the H$_2$ molecule. The yellow (blue) color represents an accumulation (depletion) of electrons. (b) The line profiles of the $\Delta\rho(z)$ (red line) and the transferred amount of charge $\Delta Q(z)$ (green line). (c) Total DOS (black line) and LDOS (red line) of antimonene adsorbed with the H$_2$ molecule. The bands colored in black and red represent antimonene and H$_2$, respectively. The black dashed lines show the Fermi level. (d) The band structure of antimonene adsorbed with the H$_2$ molecule. The black dashed lines show the Fermi level.

***H$_2$ adsorption.*** The lowest energy configuration for the H$_2$ molecule is shown in Figure 7a, where the molecule adopts a tilted configuration with the H−H bond length of 0.75 Å and is located above the Sb atom at $d = 3.56$ Å. The $E_a$ of the H$_2$ molecule on antimonene is −0.04 eV, which is almost the same as that of graphene[57] and comparable with that of InSe[54]. Considering the wide use of graphene as a hydrogen storage material due to its ability for simultaneous stable hydrogen storage and facile release,[58] our predicted H$_2$ adsorption energy for antimonene suggests that antimonene is a promising material for hydrogen storage devices.

The charge transfer analysis (Figure 7b) shows that the H$_2$ molecule is a donor to antimonene with the moderate charge transfer of 0.138 $e$ per molecule. The DCD isosurface plot (Figure 7a) indicates a depletion of electrons in both H atoms of the H$_2$ molecule and an accumulation of electrons at the nearest Sb atoms. Notably, owing to the similar presence of lone-pair electrons of the surface of phosphorene, InSe and antimonene, it is interesting to compare the charge doping behavior of H$_2$ among them. We found that the H$_2$ molecule serves as a donor for all the three cases. However, the total value of charge transfer from H$_2$ to antimonene is ten times that of H$_2$ on phosphorene while comparable to that of H$_2$ on InSe. The LDOS (Figure 7c) and the band structure (Figure 7d) analyses reveal that there are no additional H$_2$-induced states in the vicinity of the antimonene band gap. As a result, the band gap size of antimonene adsorbed with the H$_2$ molecule is the same as that of pristine antimonene (1.14 eV).

**Discussions**

**Comparison of antimonene with phosphorene and InSe**: As antimonene is a successor of phosphorene with Sb and P elements being in the same column of the periodic group, it is interesting to compare their surface chemistry with respect to the affinity to the gas molecules. It is known that one common feature of Sb and P elements is the presence of lone-pair electrons. A recent work on InSe shows that the presence of lone-pair electrons of the surface Se atoms allows the Lewis base-acid reaction with the surface species.[59] To compare these recently emerging antimonene, phosphorene and InSe 2D materials, Figure 8 plots the $E_a$-$\Delta q$ relationship for adsorption of small molecules on their surfaces. For antimonene, the $\Delta q$ is nearly linearly correlated with $E_a$ for most of the molecules with the exception of H$_2$ and NH$_3$ molecules. This indicates that the redox process associated with the charge transfer dominates the noncovalent interaction of these molecules with antimonene. The linear trend is also largely true in InSe but absent in phosphorene. Clearly, the overall slope of the $E_a$-$\Delta q$ curve of phosphorene is higher than that of antimonene. This may be due to a more electronegative nature of P than Sb. Considering that As occupies the same column in the periodic table as P and Sb, it will be interesting to examine the $E_a$-$\Delta q$ relationship of arsenene, a layered structure consisting of As atoms. All the considered molecules except the H$_2$ molecule adsorbed on antimonene lead to *p*-type doping. In comparison with phosphorene, the binding strengths of NO$_2$, O$_2$, NO and H$_2$O molecules are much stronger, while those of CO, NH$_3$ and H$_2$ molecules are weaker on antimonene. Notably, for the H$_2$ adsorption, the amount of charge transferred from antimonene is 0.138 *e* per molecule, which is ten times that of phosphorene (0.013 *e* per molecule) and comparable with that of InSe (0.146 *e* per molecule). The underlying reason might be attributed to the work function of antimonene, which is comparable to that of InSe but much higher than that of phosphorene. In contrast, the adsorption energy of H$_2$ on antimonene is much weaker than on phosphorene, largely due to the weak van der Waals interaction with the heavy Sb element.

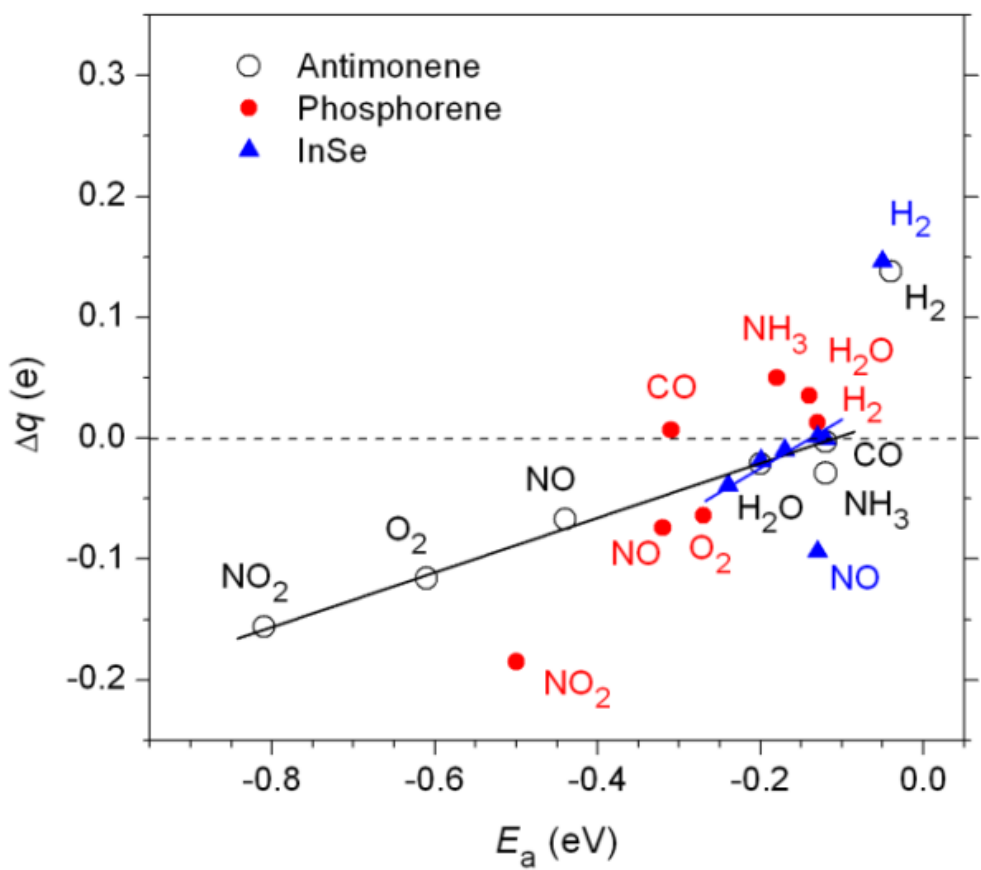

**Figure 8.** Relationships between the adsorption energy $E_a$ and charge transfer $\Delta q$ for various molecules physisorbed on antimonene, phosphorene and InSe. The results for phosphorene and InSe are adopted from Refs. 33 and 54, respectively.

**Oxidation kinetics and mechanisms of good structural stability in antimonene.** The interaction of O$_2$ molecules with 2D materials plays a critical role in their stability and performance at ambient conditions as oxidization is the most popular form of structural degradation. The interaction energy between the $O_2$ molecule and antimonene $E_a$ is found to be –0.61 eV, which is more than two times that between the $O_2$ molecule and phosphorene (–0.27 eV). A similar situation is found for the charge transfer (see Table 1 and Figure 8). These results are reasonable since Sb is less electronegative than P, which leads to a greater transfer of electrons to the $O_2$ molecule from antimonene. Our findings suggest that the performance of antimonene, for example, the carrier density and mobility, tends to be highly sensitive to the environmental $O_2$ molecule.

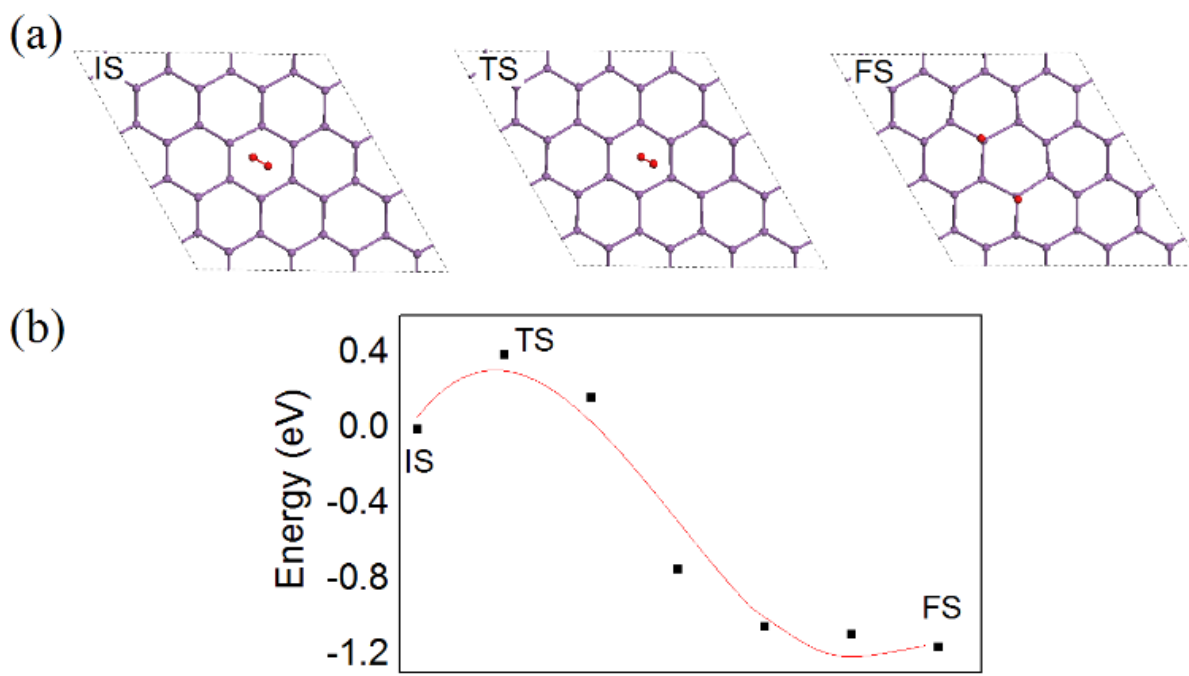

**Figure 9.** Activation barrier for the splitting of the $O_2$ molecule on antimonene: (a) the atomic models for the initial state (IS), transition state (TS) and final state (FS) state; (b) energy profile obtained by the NEB calculation for the decomposition of the $O_2$ molecule on antimonene.

To predict the oxidization behavior, thermodynamics analysis is insufficient. Therefore, we also perform kinetic analysis on the splitting of the $O_2$ molecule on the antimonene sheet in the form of terminated –O groups. The result from the climbing nudged elastic band (NEB) calculation of the above process is shown in Figure 9. Surprisingly, the energy barrier for the decomposition of the $O_2$ molecule into two apical –O groups is only ~0.40 eV. Such a small barrier implies that antimonene may undergo oxidation during synthesis and applications. Our prediction is consistent with the experimental findings that there are always some oxygen species above the surface of synthesized antimonene flakes.[11, 12] Previously, the oxidation layer containing antimonene oxide was reported to have exotic electronic properties.[60]

The predicted facile formation of oxygen species in the antimonene sheet is somehow surprising since as shown in the phosphorene case, these oxygen species tend to react with environmental $H_2O$ molecules, which leads to the degradation of the material by forming acids.[61] However, antimonene was reported to exhibit a good stability in ambient conditions.[9, 12] Hence, the roles of $O_2$ and $H_2O$ molecules and their cooperative effect on the stability of antimonene must be different from those in phosphorene. According to Table 1 and Figure 8, while the $O_2$ molecule plays the same role (acceptor) in phosphorene and antimonene, the $H_2O$ molecule behaves oppositely: it is an acceptor ($[H_2O]^{-\delta}$ with $\delta$ being a small positive real number) for antimonene but a donor ($[H_2O]^{+\delta}$) for phosphorene. Inspired by the well-known mechanism of $H_2CO_3$ acid formation from $CO_2$ and $H_2O$ molecules, which occurs through the diffusion of the $H^{+\delta}$ ion in a partially positively charged $H_2O$ to the negatively charged –O group in the $CO_2$ molecule, herein, we propose that the mechanism of the antimonene stability is related to the electrostatic repulsion between $[H_2O]^{-\delta}$ and –O group (also negatively charged -$O^{-\gamma}$ with $\gamma$ being a small positive real number). The negatively charged $[H_2O]^{-\delta}$ makes the formation and the diffusion of $H^{+\delta}$ proton to the -$O^{-\gamma}$ group unfavorable. In addition, the high stability of antimonene may also be related to the much longer Sb-Sb bond, which makes the transfer of $H^{+\delta}$ more difficult than the shorter P-P bond in phosphorene. The reason is that the transport of proton between water molecules should depend on the separation of the molecules. Owing to the larger lattice constant in antimonene, the anchored water molecules above Sb atoms should be more sparsely distributed, which reduces the hopping probability of proton among the water molecules.

The stable surface oxidation layer may be helpful for protecting the underneath antimonene layer against its degradation upon interaction with environmental molecules. Therefore, by comparing the charge transfer behavior in antimonene and InSe, we suggest that systems with $H_2O$ molecules acting as acceptor groups tends to be stable as they are less likely to form acids under the co-adsorption of $O_2$ and $H_2O$ molecules. The opposite implication is also true if $H_2O$ acts as a donor the structure tends to be decomposed: one example is easily degradable phosphorene where $H_2O$ is charge donor[33]. It should be pointed out that the presence of atomic defects, like vacancies, can change the trend of charge transfer behavior of the $O_2$ and $H_2O$ molecules and the kinetics process[32], and thus the situation may be different in the presence of defects.

**Conclusions**

By using first−principles calculations, we study the energetics and charge transfer of CO, NO, $NO_2$, $H_2O$, $O_2$, $NH_3$ and $H_2$ molecules adsorbed on antimonene. It is found that NO, $NO_2$, $H_2O$, $O_2$ and $NH_3$ molecules are effective acceptors to antimonene, while $H_2$ serves as a donor. The $NO_2$ has the strongest adsorption energy of −0.81 eV among all the considered molecules, which may arise from the coexistence of a large dipole moment of $NO_2$ and resonant molecular levels with the antimonene states. The strong acceptors like $NO_2$, NO and $O_2$ bind more strongly to the antimonene surface than the phosphorene surface, while the weak acceptors like CO, $H_2$ and $NH_3$ show a weaker adsorption.

We also examine the kinetics process of the $O_2$ molecule splitting on antimonene and find a relatively low barrier of ~0.4 eV for the $O_2$ decomposition, suggesting that antimonene tends to be oxidized during synthesis and applications largely due to the $O_2$ molecule rather than the water effect. Interestingly, the acceptor role of water impedes the interaction between water molecules and oxygen species on antimonene to form acids, which may be the underlying reason of the high stability of antimonene. The stable oxide layer may serve as a protecting layer for layers underneath it. While such oxide layer can serve as passivating and protecting layers for avoiding the degradation of layers underneath, for achieving a robust performance, potential antimonene devices still need to be protected via using nonconvalent functionalization[61] for suppressing the strong effect from environmental molecules as we predicted in this work.

**Acknowledgements**

The authors acknowledge the financial support from the Agency for Science, Technology and Research (A*STAR), Singapore, and the use of computing resources at the National Supercomputing Centre, Singapore. This work was supported in part by a grant from the Science and Engineering Research Council (152-70-00017) and the Ministry of Education, Singapore (Academic Research Fund TIER 1—RG128/14). S. V. Dmitriev acknowledges financial support from the Russian Science Foundation Grant N 14-13-00982.

**References**

1   K. Geim and K. S. Novoselov, *Nat. Mater.,* 2007, **6**, 183.

2　L. Li, Y. Yu, G. J. Ye, Q. Ge, X. Ou, H. Wu, D. Feng, X. H. Chen and Y. Zhang, *Nat. Nanotechnol.,* 2014, **9**, 372−377.

3　X. Zhang and Y. Xie, *Chem. Soc. Rev.,* 2013, **42**, 8187−8199.

4　X. Song, J. Hu and H. Zeng, *J. Mater. Chem. C*, 2013, **1**, 2952−2969.

5　Y. Yu, C. Li, Y. Liu, L. Su, Y. Zhang and L. Cao, *Sci. Rep.,* 2013, **3**, 1866.

6　S. Zhang, S. Liu, S. Huang, B. Cai, M. Xie, L. Qu, Y. Zou, Z. Hu, X. Yu and H. Zeng, *Prog. Mater. Sci.*, 2013, **58**, 1244−1315.

7　Q. Tang, Z. Zhou and Z. Chen, *WIREs Comput. Mol. Sci.*, 2015, **5**, 360−379.

8　S. Zhang, Z. Yan, Y. Li, Z. Chen and H. Zeng, *Angew. Chem., Int. Ed.*, 2015, **54**, 3112–3115.

9　P. Ares, F. Aguilar-Galindo, D. Rodriguez-San-Miguel, D. A. Aldave, S. Diaz-Tendero, M. Alcami, F. Martin, J. Gomez-Herrero and F. Zamora, *Adv. Mater.,* 2016, **28**, 6332–6336.

10　T. Lei, C. Liu, J. L. Zhao, J. M. Li, Y. P. Li, J. O. Wang, R. Wu, H. J. Qian, H. Q. Wang and K. Ibrahim, *J. Appl. Phys.*, 2016, **119**, 015302.

11　C. Gibaja, D. Rodriguez-San-Miguel, P. Ares, J. Gomez-Herrero, M. Varela, R. Gillen, J. Maultzsch, F. Hauke, A. Hirsch, G. Abellan and F. Zamora, *Angew. Chem. Int. Ed.,* 2016, **55**, 14345–14349.

12　Ji, J.; Song, X.; Liu, J.; Yan, Z.; Huo, C.; Zhang, S.; Su, M.; Liao, L.; Wang, W.; Ni, Z.; et al. Two-Dimensional Antimonene Single Crystals Grown by van der Waals Epitaxy. *Nat. Commun.*, **2016**, *7*, 13352.

13　M. Fortin-Deschênes, O. Waller, T. O. Mentes, A. Locatelli, S. Mukherjee, F. Genuzio, P. L. Levesque, A. Hébert, R. Martel and O. Moutanabbir, *Nano Lett.,* 2017, **17**, 4970.

14　X. Wu, Y. Shao, H. Liu, Z. Feng, Y. L. Wang, J. T. Sun, C. Liu, J. O. Wang, Z. L. Liu, S. Y. Zhu, Y. Q. Wang , S. X. Du , Y. G. Shi , K. Ibrahim and H. J. Gao. *Adv. Mater.,* 2017, **29**, 1605407.

15　P. Ares, J. J. Palacios, G. Abellán, J. Gómez-Herrero and F. Zamora, *Adv. Mater.,* 2018, **30**, 1703771.

16　S. Zhang, S. Guo, Z. Chen, Y. Wang, H. Gao, J. Gómez-Herrero, P. Ares, F. Zamora, Z. Zhuf and H. Zeng, *Chem. Soc. Rev*., **2018**, *DOI: 10.1039/C7CS00125H*.

17　S. Zhang, M. Xie, F. Li, Z. Yan, Y. Li, E. Kan, W. Liu, Z. Chen and H. Zeng, *Angew. Chem. Int. Ed.,* 2016, **55**, 1666–1669.

18　C. Kamal and M. Ezawa, *Phys. Rev. B,* 2015*,* **91**, 085423.

19　Z. Zhu, J. Guan and D. Tomanek, *Phys. Rev. B*, 2015, **91**, 161404.

20　Z. Zhu and D. Tomanek, *Phys. Rev. Lett.,* 2014, **112**, 176802.

21　M. Wu, H. Fu, L. Zhou, K. Yao and X. C. Zeng, *Nano Lett.,* 2015, **15**, 3557–3562.

22 H. Liu, A. T. Neal, Z. Zhu, Z. Luo, X. Xu, D. Tomanek and P. D. Ye, *ACS Nano*, **2014**, *8*, 4033–4041.

23 G. Wang, R. Pandey and S. P. Karna, *ACS Appl. Mater. Interfaces,* 2015, **7**, 11490–11496.

24 F. Mehmood and R. Pachter, *J. Appl. Phys.*, 2014, **115**, 164302.

25 A. C. Crowther, A. Ghassaei, N. Jung and L. E. Brus, *ACS Nano,* 2012, **6**, 1865−1875.

26 O. Leenaerts, B. Partoens and F. M. Peeters, *Phys. Rev. B: Condens. Matter Mater. Phys.,* 2008, **77**, 125416.

27 L. F. Yang, Y. Song, W. B. Mi and X. C. Wang, *Appl. Phys. Lett.,* 2016, **109**, 022103.

28 L. Yang, Y. Song, W. Mi and X. Wang, *RSC Adv.*, 2016, **6**, 66140–66146.

29 J. Gu, Z. Du, C. Zhang, J. Ma, B. Li and S. Yang, *Adv. Energy Mater.*, 2017, **7**, 1700447.

30 E. Martínez-Periñán, M. P. Down, C. Gibaja, E. Lorenzo, F. Zamora and C. E. Banks, *Adv. Energy Mater.*, 2017, **1702606**, 1–7.

31 Q. Y. He, Z. Y. Zeng, Z. Y. Yin, H. Li, S. X. Wu, X. Huang and H. Zhang, *Small*, 2012, **8**, 2994−2999.

32 A. A. Kistanov, Y. Cai, K. Zhou, S. V. Dmitriev and Y. W. Zhang, *2D Mater.*, 2017, **4**, 015010.

33 Y. Cai, Q. Ke, G. Zhang and Y. W. Zhang, *J. Phys. Chem. C,* 2015, **119**, 3102−3110.

34 J. Gao, G. Zhang and Y. W. Zhang, *Nanoscale,* 2017, **9**, 4219−4226.

35 Y. Guo, S. Zhou, Y. Bai and J. Zhao, *ACS Appl. Mater. Interfaces,* 2017, **9**, 12013−12020.

36 N. Liu and S. Zhou, *Nanotechnology*, 2017, **28**, 175708.

37 L. Chen, L. Wang, Z. Shuai and D. Beljonne, *J. Phys. Chem. Lett.*, 2013, **4**, 2158−2165.

38 D. C. Elias, R. R. Nair, T. M. G. Mohiuddin, S. V. Morozov, P. Blake, M. P. Halsall, A. C. Ferrari, D. W. Boukhvalov, M. I. Katsnelson, A. K. Geim and K. S. Novoselov, *Science,* 2009, **323**, 610.

39 Y. Cai, Z. Bai, H. Pan, Y. P. Feng, B. I. Yakobson and Zhang, Y. W. *Nanoscale*, 2014, **6**, 1691−1697.

40 Y. Jing, Q. Tang, P. He, Z. Zhou and P. Shen, *Nanotechnology*, 2015, **26**, 09520.

41 X. Peng, Q. Wei and A. Copple, *Phys. Rev. B*, 2014, **90**, 085402.

42 H. Y. Lv, W. J. Lu, D. F. Shao and Y. P. Sun, *Phys. Rev. B*, 2014, **90**, 085433.

43 A. J. Samuels and J. D. Carey, *ACS Nano*, 2013, **7**, 2790–2799.

44 J. Xiao, M. Long, X. Li, Q. Zhang, H. Xu and K. S. Chan, *J. Phys.: Condens. Matter*, 2014, **26**, 405302.

45 Z. Chen, P. Darancet, L. Wang, A. C. Crowther, Y. Gao, C. R. Dean, T. Taniguchi, K. Watanabe, J. Hone, C. A. Marianetti and L. E. Brus, *ACS Nano*, 2014, **8**, 2943−2950.

46 D. J. Late, Y. K. Huang, B. Liu, J. Acharya, S. N. Shirodkar, J. Luo, A. Yan, D. Charles, U. V. Waghmare, V. P. Dravid and C. N. R. Rao, *ACS Nano*, 2013, **7**, 4879−4891.

47 B. Chen, H. Sahin, A. Suslu, L. Ding, M. I. Bertoni, F. M. Peeters and S. Tongay, *ACS Nano*, 2015, **9**, 5326–5332.

48 S. Cui, H. Pu, S. A. Wells, Z. Wen, S. Mao, J. Chang, M. C. Hersam and J. Chen, *Nat. Commun.,* 2015, **6**, 8632.

49 A. N. Abbas, B. Liu, L. Chen, Y. Ma, S. Cong, N. Aroonyadet, M. Köpf, T. Nilges and C. Zhou, 2015*,* **9**, 5618–5624.

50 G. Kresse and J. Furthmüller, *Phys. Rev. B: Condens. Matter Mater. Phys.*, 1996, **54**, 11169.

51 Becke, A. D. *Phys. Rev. A: At., Mol., Opt. Phys.*, 1988, **38**, 3098.

52 Ü. O. Aktürk, E. Aktürk and S. Ciraci, *Phys. Rev. B*, 2016, **93**, 035450.

53 Ü. O. Aktürk, O. V. Özçelik and S. Ciraci, *Phys. Rev. B*, 2015, **91**, 235446.

54 Y. Cai, G. Zhang and Y. W. Zhang, *J. Phys. Chem. C*, 2017, **121**, 10182−10193.

55 L. Kou, T. Frauenheim and C. Chen*, J. Phys. Chem. Lett.*, 2014, **5**, 2675–2681.

56 A. A. Kistanov, Y. Cai, Y. W. Zhang, S. V. Dmitriev and K. Zhou, *J. Phys.: Condens. Matter*, 2017, **29**, 095302.

57 S. C. Wang, L. Senbetu and C. Woo, J. *Low Temp. Phys*., 1980, **41**, 611.

58 C. Ataca, E. Aktürk, S. Ciraci and H. Ustunel, *Appl. Phys. Lett.*, 2008, **93**, 043123.

59 S. Lei, X. Wang, B. Li, J. Kang, Y. He, A. George, L. Ge, Y. Gong, P. Dong and Z. Jin, *Nat. Nanotechnol.*, 2016, **11**, 465.

60 S. Zhang, W. Zhou, Y. Ma, J. Ji, B. Cai, S. A. Yan, Z. Zhu, Z. Chen and H. Zeng, *Nano Lett.*, 2017, **17**, 3434–3440.

61 Y. Huang, J. Qiao, K. He, S. Bliznakov, E. Sutter, X. Chen, D. Luo, F. Meng, D. Su, J. Decker, W. Ji, R. S. Ruoff and P. Sutter, *Chem. Mater.*, 2016, **28**, 8330–8339.

62 G. Abellán, P. Ares, S. Wild, E. Nuin, C. Neiss, D. R. Miguel, P. Segovia, C. Gibaja, E. G. Michel, A. Görling, F. Hauke, J. Gómez-Herrero, A. Hirsch and F. Zamora, *Angew. Chem. Int. Ed.*, *2017*, **56**, 14389−14394.